\documentclass[aps,prl,preprint,groupaddress,showpacs]{revtex4}
\usepackage{graphicx}
\usepackage{amsmath}
\usepackage{amsfonts}
\usepackage{amssymb}
\usepackage{dsfont}
\usepackage{dcolumn}
\usepackage{bm}

\newcommand{\beqn}{\begin{eqnarray}}
\newcommand{\eeqn}{\end{eqnarray}}
\newcommand{\be}{\begin{equation}}
\newcommand{\ee}{\end{equation}}

\newcommand{\mathsym}[1]{{}}

\def \cha{\widetilde{\chi}^{\pm}_1}

\def \ta{\widetilde{t}_1}
\def \tb{\widetilde{t}_2}

\def \sta{\widetilde{\tau}_1}

\begin{document}

\title{  Landscape of Supersymmetric Particle Mass Hierarchies and their Signature Space at the CERN Large Hadron Collider
 }

\author{  Daniel Feldman,   Zuowei Liu and   Pran Nath}

\affiliation{Department of Physics, Northeastern University,
 Boston, MA 02115, USA \\ Published in :  {\rm Phys. Rev. Lett. 99, 251802 (2007)}
  \\(Received 13 July 2007; Published 18 December 2007)
  }
\begin{abstract}
The minimal supersymmetric standard model with soft breaking has a large landscape of supersymmetric particle mass hierarchies. This number is reduced significantly in well-motivated scenarios
such as minimal supergravity and alternatives. We carry out an analysis of the landscape for the first four
lightest particles and identify at least 16 mass patterns, and provide benchmarks for each.   We study the
signature space for the patterns at the CERN Large Hadron Collider by analyzing the lepton + (jet
$\geq 2$) + missing $P_T$  signals with 0, 1, 2 and 3  leptons. Correlations in missing $P_T$ are also analyzed. It is found that even with 10 fb$^{-1}$ of data a significant discrimination among patterns emerges.
 \end{abstract}
 \maketitle
{\it Introduction}:
The search for supersymmetric particles (sparticles)  is one of the major goals of the current experiments at the Fermilab
Tevatron collider and
at the CERN Large  Hadron Collider (LHC)  which will  come on line
in the very  near future.
Central to the discovery of sparticles is the way their masses align in a hierarchical mass
pattern as such alignment has strong influences on the ability of experiments  to probe signatures
emerging from sparticle production at colliders.
There are 32 supersymmetric masses  in  the minimal supersymmetric standard model
 (MSSM)
including the Higgs bosons. With the sum rule constraints in the
gaugino and Higgs sectors but without any phenomenological
constraints the number of mass  parameters is 27. Using Stirling's
formula [$n!\sim \sqrt{2\pi n} (n/e)^n$] one finds at least
$O(10^{28})$ possibilities for hierarchical patterns. Although this
number does not  rise  to the level  of $O(10^{1000})$ as in string
landscapes, it is still an impressive number and can be construed as
a mini landscape in this context. The number of possibilities  is
drastically reduced in the minimal supergravity grand unified model
mSUGRA \cite{msugra,bfs} although no classification has ever been made
and the precise  number of possibilities is not known.
In this Letter we undertake this cartography. To keep the analysis under control we focus on
mass hierarchies for the first four sparticles excluding the
lightest Higgs. Such a cartography is important for devising
strategies for analyzing data from the LHC.

Our analysis, based on a large scan of the mSUGRA parameter
space, reveals at least 16 hierarchical patterns for the first four
sparticles, of which a significant number do not appear in
 conventional benchmarks such
as Snowmass points \cite{Allanach:2002nj}
(these are a representative sample of points in the MSSM parameter space  called benchmarks which are used for theoretical predictions of supersymmetry at present and future colliders)
and Post WMAP
benchmarks \cite{Battaglia:2003ab} (which is an updated version of previous benchmarks taking
account of the relic density constraints from data from the Wilkinson Microwave Anisotropy Probe (WMAP)).
\begin{figure}[h]
\includegraphics[width=9.3cm,height=9.3cm]{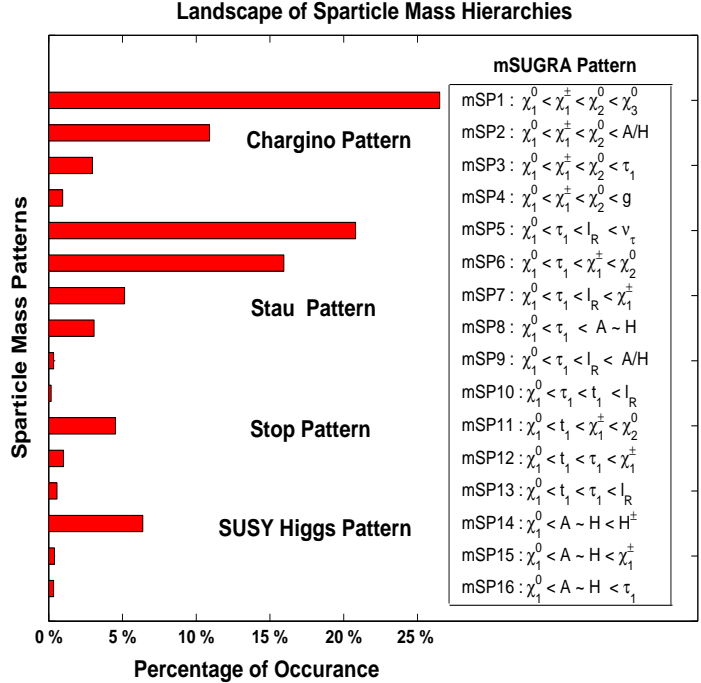}
\caption{The surviving hierarchical patterns in the mSUGRA landscape
with constraints as discussed in the text.  In mSP14,15,16  the
$\tilde \chi_1^0$  and the Higgs bosons $(A,H)$ are sometimes seen to
switch.} \label{landscape}
\end{figure}
The analysis is carried  out in the mSUGRA model (for recent works see
\cite{Djouadi:2006be,Arnowitt:2007nt,Ellis:2006ix}).
mSUGRA is an
effective theory below the Grand Unification scale $M_G$
and its parameter space is defined by the parameters
 $( m_0, m_{1/2}, A_0, \tan\beta, {\rm sign} \mu)$
where  $(m_0,m_{1/2})$ are the universal scalar and gaugino masses,
$A_0$ is the universal trilinear coupling, and $\tan \beta$
is the  ratio of the vacuum expectation values (VEVs) of the two neutral Higgs fields
in MSSM, and sign$\mu$ is  the sign of  $\mu$, where $\mu$ is the Higgs
mixing parameter.
In the analysis we impose
the relic density constraints  on the abundance of the lightest
neutralino ($\tilde \chi_1^0$)
consistent with the WMAP: $ 0.0855<\Omega_{\tilde \chi_1^0}
h^2<0.1189$ \cite{Spergel:2006hy}, and other experimental
constraints as follows: $2.83\times10^{-4}< \rm{BR}(b\rightarrow
s\gamma)<4.63\times 10^{-4}$ (where the SUSY loop contributions to
this decay can be comparable to the corrections in the Standard Model (SM)), $m_h>
100 ~{\rm GeV}$, $m_{\cha}>104.5 ~{\rm GeV}$, $m_{\ta}>101.5 ~{\rm
GeV}$, $m_{\sta}>98.8 ~{\rm  GeV}$,
where $h,\tilde  \chi_1^{\pm}, \tilde t_1, \tilde \tau_1$ are the lightest Higgs boson,
chargino, stop and stau respectively.

For the calculations of the
relic density of $\tilde \chi_1^0$  we use MicrOMEGAs version 2.0.1
\cite{MICRO} with sparticle  and Higgs masses calculated using the
package SUSPECT 2.3
of Ref. \cite{SUSPECT}.  We have investigated other softwares
\cite{ISAJET,SPHENO,SOFTSUSY,Baer:2005pv,Belanger:2005jk,Allanach:2004rh,Allanach:2003jw}
and find significant agreement among them. For each mSUGRA
model point  that survives the constraints mentioned above, we
compute the sparticle
\begin{figure*}[t]
\includegraphics[width=7.8cm,height=6.2cm]{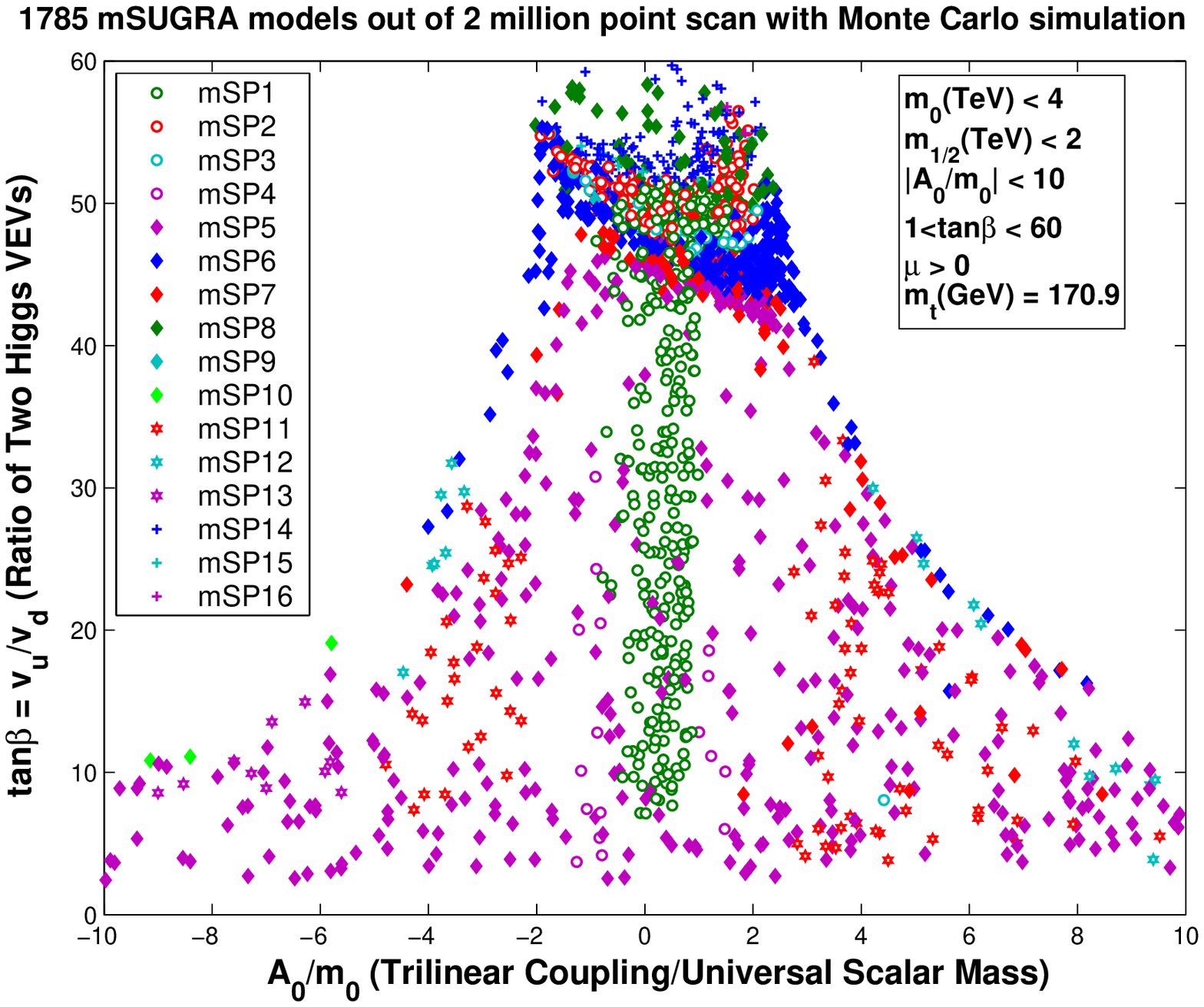}
\hspace*{.2in}\includegraphics[width=8cm,height=6.4cm]{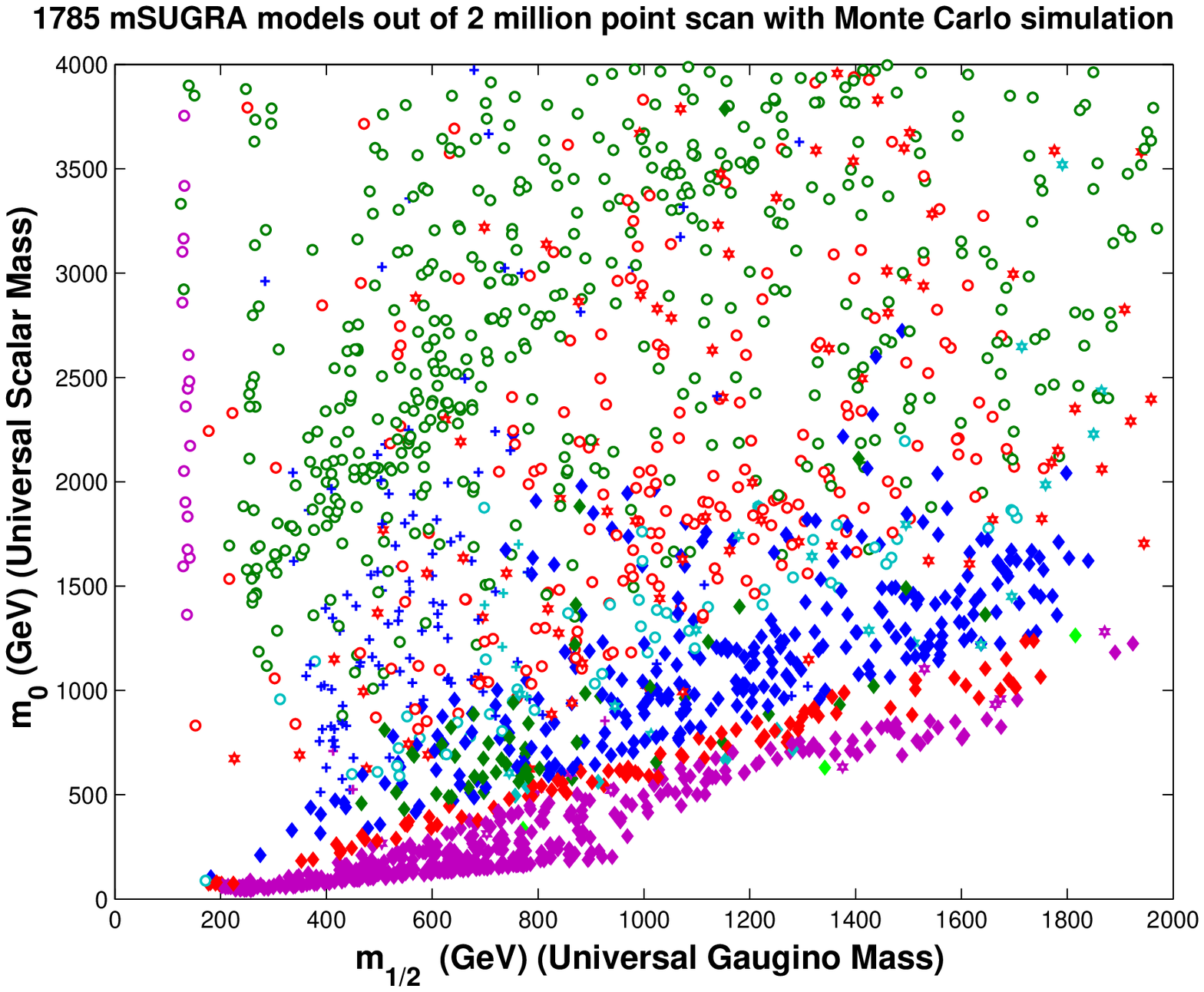}
\caption{The plots
exhibit the 16 mass patterns (see Fig.(1))  in $\tan\beta-A_0/m_0$  and  $m_0-m_{{1}/{2}}$ planes
among the 1785 surviving
points out of a
random scan using $2\times 10^6$ models with flat priors in the
ranges indicated in the left figure panel consistent with the
constraints discussed in the text.}
\label{2m}
\end{figure*}
spectrum, mixing angles, etc.  The resultant SUSY Les Houches
Accord (SLHA) \cite{SKANDS} file is called by the
PGS4
olympics main
Fortran file, which simulates the LHC detector effects and event
reconstruction (see \cite{PGS} for detailed discussion). For the
computation of SUSY production cross sections and branching
fractions we employ PYTHIA 6.4.11 \cite{PYTHIA}, which simulates
hadronization and event generation from the fundamental SUSY
Lagrangian. SUSY cross section and events were  generated using all
32 sparticles in this analysis. We have cross checked
sample points with PROSPINO \cite{PROSPINO} which computes the next-to-leading
order cross sections for the production of supersymmetric particles at hadron colliders,
and TAUOLA \cite{TAUOLA}  is
called by PGS4 for the calculation of tau branching fractions. With
PGS4 we use the Level 1
 (L1) triggers based on  the Compact Muon Solenoid detector
 (CMS) specifications \cite{CMS} and the LHC detector card.

{\it Landscape of mass hierarchies}: We focus on the patterns for
the four lightest particles (discounting the lightest Higgs) as they
would to a great degree influence the discovery of SUSY while
keeping the size of the landscape in check. We have carried out a
mapping of the mSUGRA parameter space with $2\times 10^6$ models
with $\mu > 0$ and the parameter range given in Fig.(\ref{2m}) via
Monte Carlo scan with flat priors under the constraint of radiative
breaking of the electroweak symmetry. 55 hierarchical patterns for
the first four particles were seen which are reduced to 16 with
relic density and collider constraints. We label these as  mSUGRA
pattern 1 (mSP1) through mSUGRA pattern 16 (mSP16)  as shown in
Fig.(\ref{2m}). The frequency of their  occurrence is exhibited  in
Fig.(\ref{landscape}).  A significant set of the mSP1 points lie
in the  region $|A_0/m_0|<1$ and correspond to the Hyperbolic
Branch/ Focus Point (HB/FP) region \cite{hb}. In Table \ref{pattern} we also give
illustrative, mass pattern motivated, benchmark points, for  each of
the mass patterns mSP1-mSP16 chosen with a moderate to light SUSY
scale ($Q =\sqrt{m_{\ta}m_{ \tb}}$) and to show the effects of
scanning over the parameter space.
\begin{table}[t]
\begin{center}
\caption{Benchmarks  using SUSPECT 2.3  with one point for each
mass pattern mSP1-mSP16. Also given are the neutralino LSP
(Lightest SUSY (R parity odd) Particle), and the Lightest Charged
Particle (LCP) masses. We take ${\mu}
> 0$, ${m_b}^{\overline{\rm MS}}(m_b) = 4.23$ {\rm GeV},
${\alpha_s}^{\overline{\rm MS}}(M_Z)=.1172$, and $m_t({\rm pole}) =
170.9$ ${\rm GeV}$. At least five LCP from these benchmarks will be
accessible  at the International  Linear Collider (ILC).}
\begin{ruledtabular}
\begin{tabular}{c|c |c |c |c| c |c c   }
mSUGRA  &    $m_0$   &   $m_{1/2}$   &   $A_0$ &  $\tan\beta$  & $\mu(Q)$  &   LSP$~|~$LCP             \\
Pattern &     $({\rm GeV})$   &   $({\rm GeV})$ & $ ({\rm GeV})$     &  $v_u/v_d$  &   $({\rm GeV})$    &   $({\rm GeV})$        \\
\hline
mSP1:   &     2001&411&0&30     & 216     &156.1$~|~$202.6     \\
mSP2:   &    1125&614&2000&50   & 673      &256.7$~|~$483.1  \\
mSP3:   &    741&551&0&50        & 632    &230.5$~|~$434.7   \\
mSP4:   &     1674&137&1985&18.6 & 533   &54.3$~|~$106.9\\\hline
mSP5:   &     111&531&0&5       & 679     &217.9$~|~$226.3  \\
mSP6:   &      245&370&945&31    & 427    &148.6$~|~$156.8  \\
mSP7:   &      75 &201& 230& 14   & 246    &74.8$~|~$100.2 \\
mSP8:   &      1880&877&4075&54.8 & 1141     &373.1$~|~$379.6 \\
mSP9:   &     667& 1154&-125&51   & 1257     &499.2$~|~$501.8  \\
mSP10:  &      336&772&-3074&10.8  & 1695 &329.2$~|~$331.7\\\hline
mSP11:  &      871&1031&-4355 &10 & 2306 &447.1$~|~$491.5 \\
mSP12:  &      1371&1671&-6855&10 & 3593 &741.2$~|~$791.8  \\
mSP13:  &     524&800&-3315 & 15 & 1782 &342.7$~|~$383.8  \\\hline
mSP14:  &     1036&562&500 & 53.5 & 560&236.2$~|~$399.1  \\
mSP15:  &       1113&758&1097&51.6 & 724 &321.1$~|~$595.9  \\
mSP16:  &       525&450&641&56     & 484 &184.6$~|~$257.9  \\
\end{tabular}
\end{ruledtabular}
\label{pattern}
\end{center}
 \end{table}
\begin{figure*}[t]
\centering
\hspace*{-.1in}\includegraphics[width=8cm,height=6.0cm]{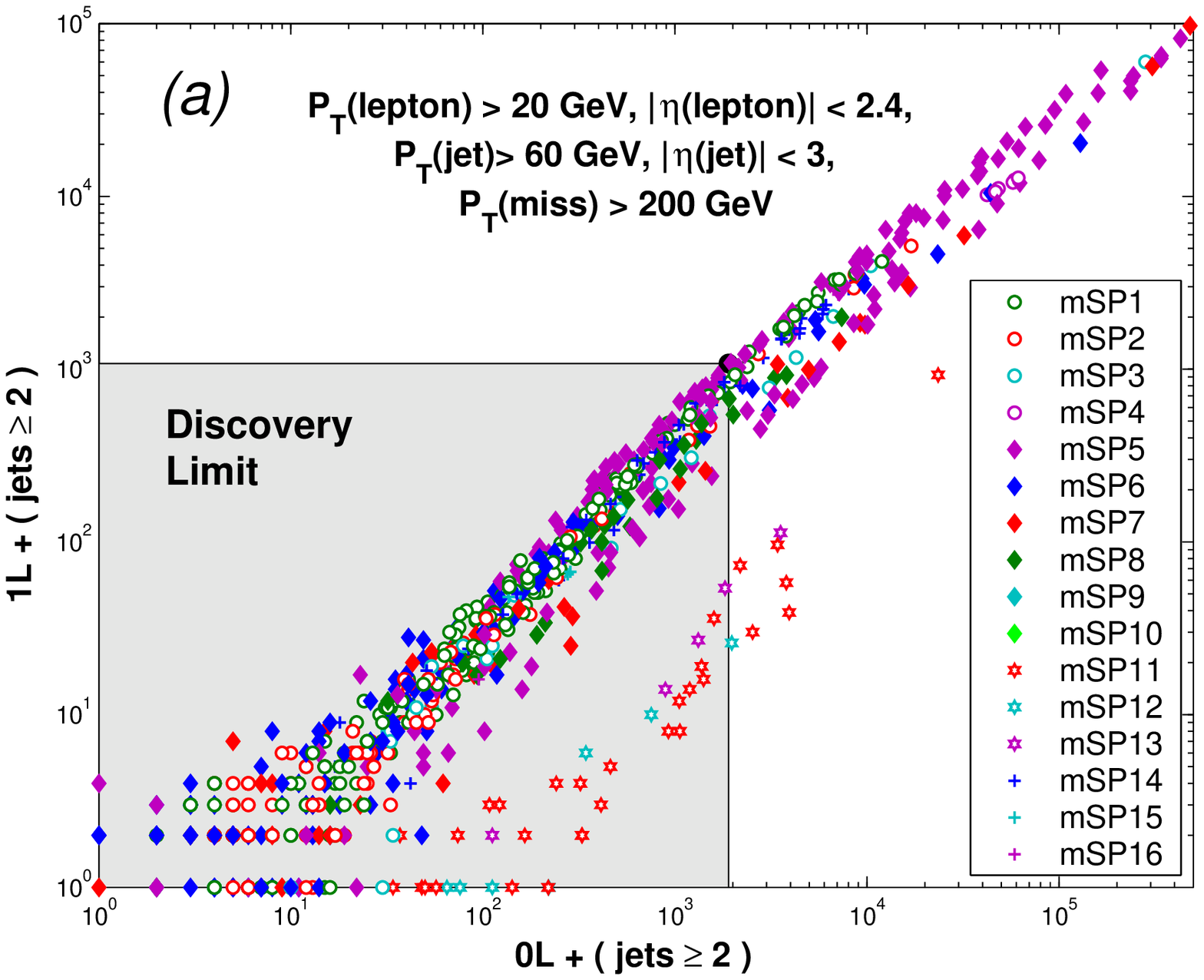}
\hspace*{.2in}\includegraphics[width=8cm,height=6.0cm]{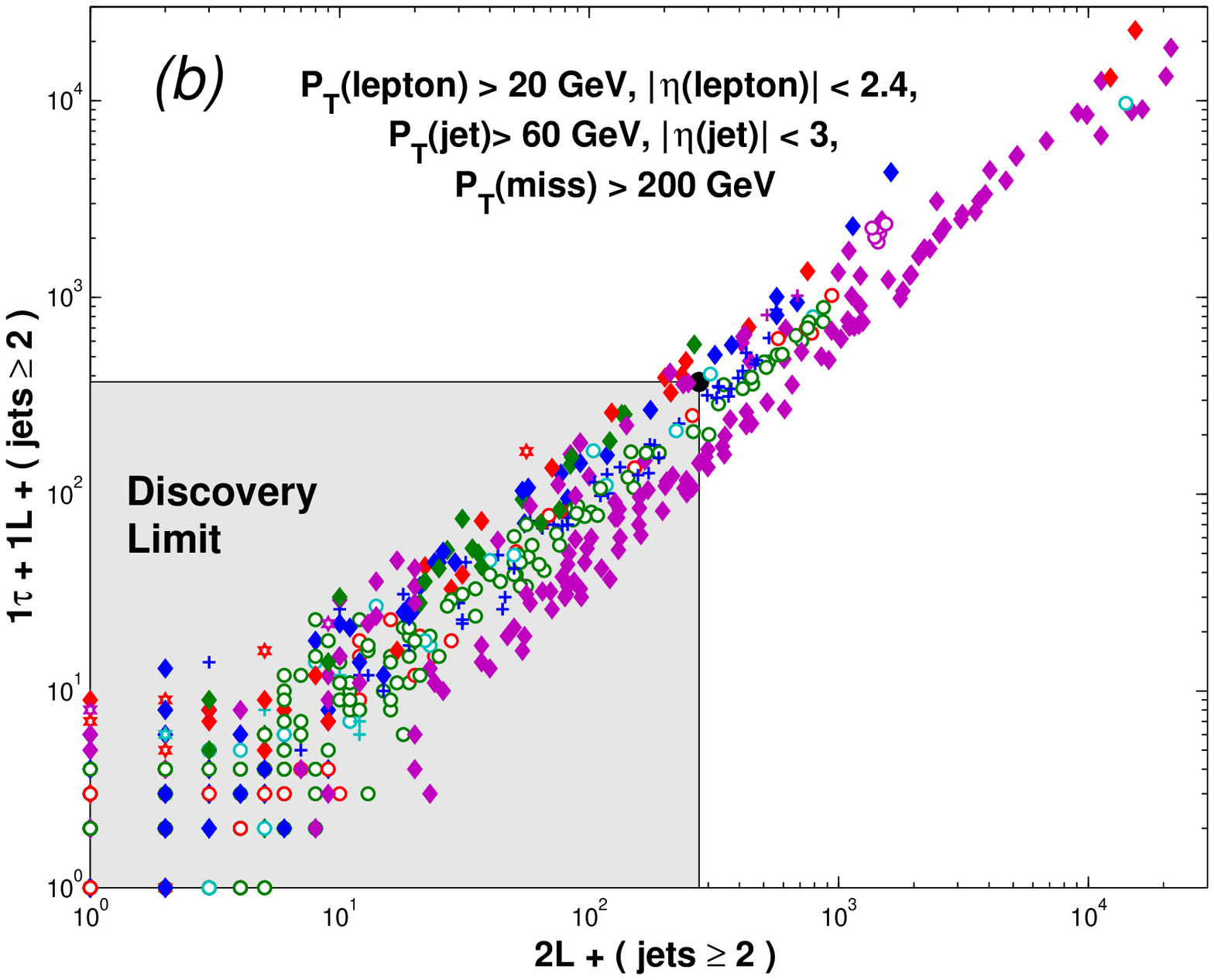}
\caption{Leptons + $n$ jets ($n\geq 2$)  signatures originating from 16 patterns with
$P_T$ and rapidity $\eta$ cuts as shown where
 (a) single-lepton vs no-lepton; (b) single-lepton plus single-$\tau$ vs dilepton.
the shaded regions in (a)-(b) are due to SM backgrounds
$t \bar{t}, b \bar{b}$, Dijets, Drell-Yan, and $Z,W$ production.
The discovery limits are max$\{5\sqrt{N_{SM}}, 10\}$, where $N_{SM}$ is the Standard
Model background.
}
 \label{set1}
\end{figure*}
\begin{figure*}[t]
  \centering
\hspace*{-.1in}\includegraphics[width=8cm,height=6.0cm]{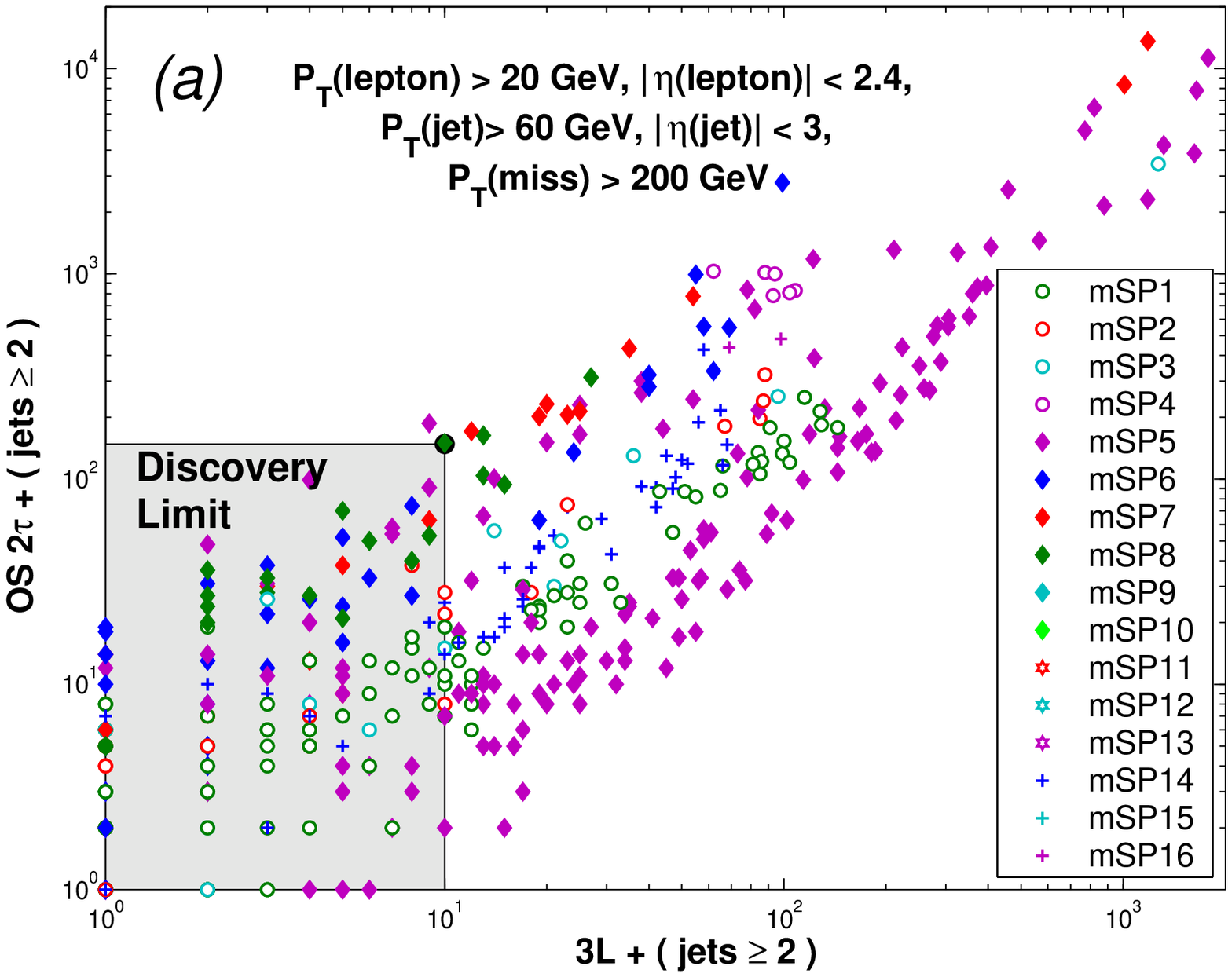}
\hspace*{.2in}\includegraphics[width=8cm,height=6.0 cm]{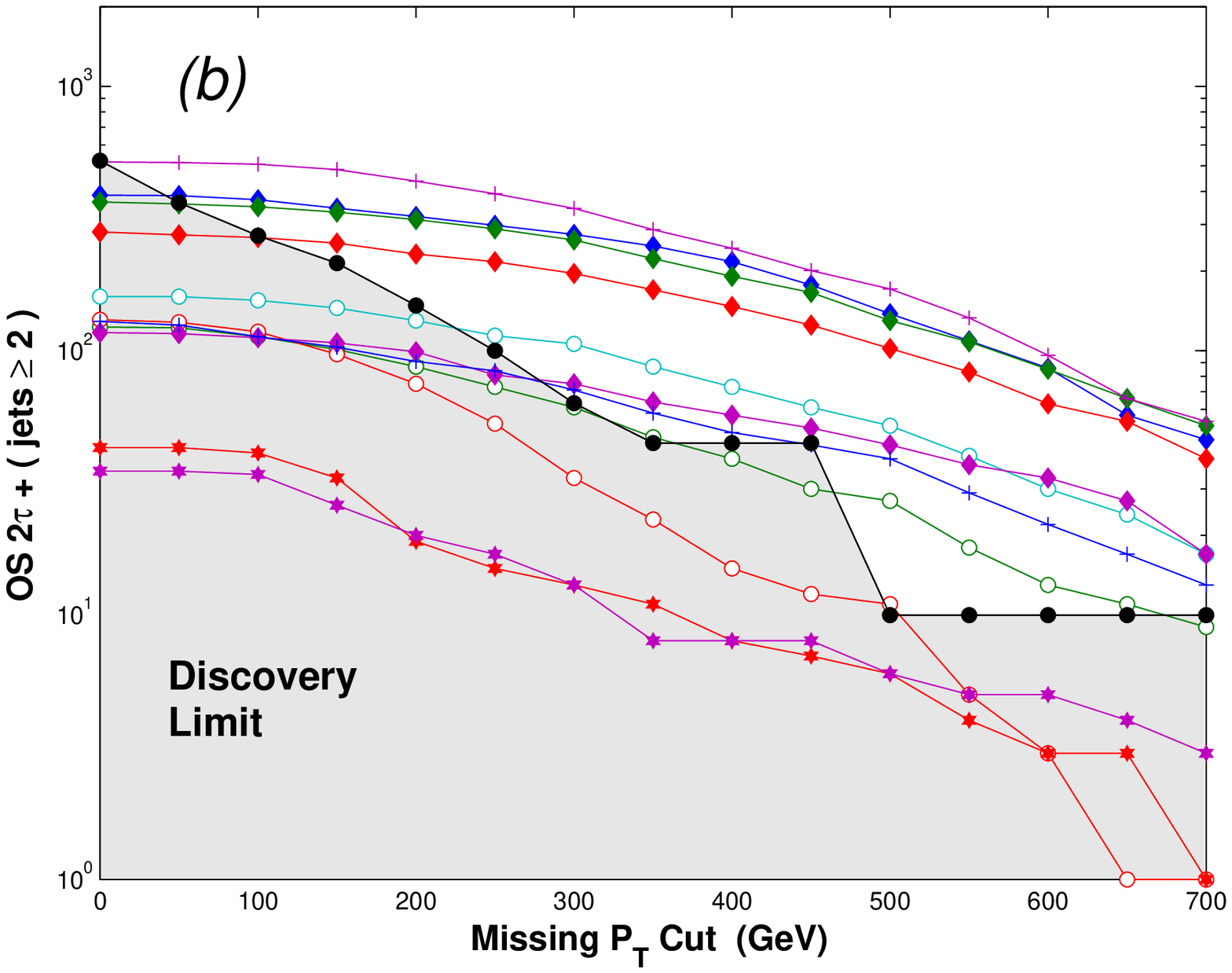}
\caption{The number of $\tau^{+}\tau^{-}$ events vs (a) trilepton
events, and vs (b) $P_T^{miss}$  cut. In (b) the
 post
trigger level cuts are as in Figs.(\ref{set1}) and (4a) except for  the
$P_T^{miss}$  cut.
Each curve is  for one point for each mSP, chosen
such that their PGS4 data files are nearly of the same (5MB) size.}
 \label{set2}
\end{figure*}
Most of  the mSP patterns do not appear in previous works. Thus all
the Snowmass mSUGRA points (labeled SPS) \cite{Allanach:2002nj}
 are only of types mSP(1,3,5,7) as follows:
 ${\rm (SPS1a, SPS1b, SPS5)}$ $\to$  ${\rm mSP7}$,
 ${\rm SPS2}$ $\to$  ${\rm mSP1}$,  ${\rm SPS3}$ $\to$ ${\rm mSP5}$, and
 ${\rm (SPS4,SPS6)}$ $\to$   ${\rm mSP3}$.
Regarding the Post-WMAP points \cite{Battaglia:2003ab} one has the
following  mapping ${\rm (A', B', C', D', G', H', J',M')}$ $\to$
{\rm mSP5}, $({\rm I',L'})\to {\rm mSP7}$, ${\rm E'}\to {\rm mSP1}$,
${\rm K'}\to {\rm mSP6}$.
We observe more patterns owing to the
sampling of larger regions of the  parameter space especially in $A_0$.

{\it Mass hierarchies  and their  signatures}: We have carried out
analyses of the  lepton + jet signals to determine if such signals
can discriminate among the patterns. We have analyzed 0, 1, 2 and 3
leptonic events (lepton = L, $\tau$; L = e, $\mu$) with and without
jets.
 Our Standard Model backgrounds are checked
against two CMS notes \cite{CMSnotes} and our results lie within the
error bars of these analyses.  Some  analyses of this type
exist\cite{chameleon,Conlon:2007xv,Kane:2006yi,Baer:2007ya} but not
in the context of mass hierarchies  analyzed here. We give now the
details of the analysis.   In Fig.(\ref{set1}), we use 902 sample
models from Fig.(\ref{2m}) (corresponding to a scan of $10^6$ points
as it contains all the essential
features of the larger scan and is visually clearer)
to generate SUSY signals through PGS4
using 10 fb$^{-1}$ of integrated luminosity. In Fig.(\ref{set1}a) an
analysis of the  ${\rm 1L}+ ({\rm jets}\geq 2)$  vs  ${\rm 0L}+
({\rm jets}\geq 2)$ events is given along with the discovery limit
for each of these model points.
 Here one is beginning to see discrimination among patterns. Specifically the signals of
mSP11-mSP13  lie significantly lower than the others.
 In  Fig.(\ref{set1}b) we give  an analysis of
 $1\tau+{\rm 1L}+ ({\rm jets}\geq 2)$  vs   ${\rm 2L}+ ({\rm jets}\geq 2)$.
We see  that relative  to Fig.(\ref{set1}a) there is now more
dispersion among the patterns and furthermore the SM  background is
also  significantly smaller in this case.  Finally in
Fig.(\ref{set2}a) we have  a plot of ${\rm OS(Opposite~Sign)}2\tau+({\rm jets}\geq 2) $ vs  $3{\rm L}+ ({\rm jets} \geq
2)$. Here we see that the dispersion among models is even larger and
the SM background is even smaller. Thus we can see visually the
effect of the SM background shrinking, the signal relative to the background getting stronger
and the discrimination among  the  patterns increasing as we move
from Fig.(\ref{set1}a)  to Fig.(\ref{set2}a). As expected the
trileptonic signal \cite{Nath:1987sw}
 is the strongest with the least background.
The missing $P_T$  cuts  can also help discriminate among models.
In  Fig.(\ref{set2}b)  we give an analysis of
OS $2\tau+ ({\rm jets}\geq 2)$ vs  missing
$P_T$ cut. One  may note the
 very significant dispersion among the patterns with  missing $P_T$ cut.


{\it Conclusion}:
The
discovery of supersymmetry is one of the major goals of the
current effort at the Tevatron and   in experiments with the CMS and the ATLAS
detectors in the very near future at the LHC.
When  sparticles are produced the signatures of  their production
 will be determined by their hierarchical mass patterns.
In this Letter we have investigated hierarchical mass patterns for the four lightest
sparticles within one of the leading candidate theories - the mSUGRA model.
The analysis  shows  16 such patterns consistent with all the current
experimental constraints and most of these patterns are new and not discussed in the
previous literature.
We also carried out an  analysis of signatures of these patterns
which include (0,1,2,3) leptons + $n$ jets
($n\geq 2$) signals and also  missing transverse momentum ($P_T$) signals.
 We conclude that even with  10 fb$^{-1}$ of luminosity significant dispersion can be seen among many  mSP signatures  and thus discrimination among patterns is possible.
Similar analyses are desirable  for other plausible soft breaking scenarios beyond mSUGRA.

{\it Acknowledgements}: We thank
Brent Nelson and George Alverson
 for discussions. This work is supported in part by NSF
grant  PHY-0456568.

\end{document}